\documentclass[lettersize,journal]{IEEEtran}

\usepackage{amsfonts}
\usepackage{cite}
\usepackage{amsmath}
\usepackage{amssymb}
\usepackage{graphicx}
\usepackage{textcomp}
\usepackage{url}
\usepackage{balance}

\begin{document}

\markboth{1571281384}{1571281384}

\title{Dynamical Reduction of Two Series Josephson Junctions to a Synthetic High-Transparency Josephson Element}

\author{
Claudio Guarcello,
Sergio Pagano,
Carlo Barone,
Alessandro Bruno,
A. Mert Bozkurt,
and Giovanni Filatrella
%
%
%
%
\thanks{
Received xx yy zzzz; revised xx yy zzzz; accepted xx yy zzzz. Date of publication xx yy zzzz; date of current version xx yy zzzz.
This work was supported by Italian INFN under Grant QUARTET, and by the University of Salerno Italy under Grants FRB23BARON, FRB24CAVAL and FRB25PAGAN.
}
\thanks{
C. Guarcello, C. Barone, and S. Pagano are with 
Department of Physics ``E.~R.~Caianiello'',
University of Salerno,
I-84084 Fisciano, Salerno, Italy;
INFN, Sezione di Napoli, Gruppo Collegato di Salerno,
Complesso Universitario di Monte S. Angelo,
I-80126 Napoli, Italy;
and CNR-SPIN,
c/o University of Salerno,
via Giovanni Paolo II 132,
I-84084 Fisciano, Salerno, Italy.
}
\thanks{
A. M. Bozkurt and A. Bruno are with 
QuantWare,
Molengraaffsingel 8,
2629 JD Delft, The Netherlands.
}
\thanks{
G. Filatrella is with 
Department of Sciences and Technologies,
University of Sannio,
via de Sanctis,
I-82100 Benevento, Italy.
}
\thanks{
Corresponding author: Claudio Guarcello (e-mail: cguarcello@unisa.it). All authors contributed equally to this work.
}
}

\maketitle

\begin{abstract}
Two conventional Josephson junctions connected in series can reproduce, in the static limit in which the currents through the capacitive and resistive channels are negligible, the current-phase relation of a single effective weak link with tunable transparency. 
Therefore, the two-junction series can be treated as a single synthetic high-transparency element. 
Here, we investigate to what extent this mapping remains valid under finite-frequency drive and retaining the junctions' resistive and capacitive terms. 
The full resistively and capacitively shunted junction equations are compared with an effective synthetic element with tunable transparency that retains the synthetic tunable-transparency current-phase relation together with effective capacitive and dissipative terms, thus reducing the two second order degree of freedom system to a single second order degree of freedom. 
The resulting single-element dynamics is compared with the complete two-junction system under ac excitation. The agreement is quantified through a normalized root-mean-square error between the full and effective voltage waveforms.
A broad low-error region is found at low drive frequency, while pronounced deviations emerge as the drive frequency approaches the relevant plasma-frequency scale and at larger drive amplitudes. 
The results provide a quantitative dynamical criterion for using the reduced single-element description of a synthetic high-transparency Josephson element in superconducting circuits.
\end{abstract}


\section{Introduction}

\IEEEPARstart{T}{he} current-phase relation (CPR) of a Josephson
junction (JJ) determines its nonlinear electromagnetic response and is
a central design parameter in superconducting electronics.
Beyond the conventional sinusoidal tunnel-junction limit,
nonsinusoidal CPRs arise naturally in weak links with finite channel
transparency and contain higher Josephson harmonics that can strongly
modify both static and dynamical properties
\cite{Golubov2004,Beenakker1991,Sochnikov2015}. 
The recent observation of sizable higher harmonics even in nominally standard tunnel junctions has further emphasized that nonsinusoidal CPR can be relevant
in realistic superconducting circuits~\cite{Willsch2024}. 
More generally, tailoring Josephson energy-phase relations has become an
increasingly useful strategy for engineering selected nonlinearities
in superconducting devices, including three-wave-mixing elements and
Kerr-controlled parametric circuits
\cite{Frattini2017,Sivak2019,Zorin2016,Ranadive2022,Buccheri2026}.

This has motivated growing interest in engineering elements with tailored
CPRs by synthesizing nonsinusoidal Josephson responses from conventional
junctions. Bozkurt \emph{et al.} showed that two conventional, i.e., with a sinusoidal CPR, JJs in series
can reproduce the energy-phase relation of a short single-channel weak
link, with an effective transparency controlled by the junction asymmetry
\cite{Bozkurt2023}. This concept was subsequently demonstrated in
voltage-controlled hybrid Josephson circuits~\cite{Banszerus2024} and
extended to hybrid Josephson rhombi with tunable $\cos(2\varphi)$
responses and superconducting-diode regimes~\cite{Banszerus2025}.
Related multi-junction architectures have been explored for
$\cos(2\varphi)$ qubits and in comparison with Andreev weak links
\cite{Zhurbina2026,Bozkurt2023proceeding}. Such synthetic CPR engineering
provides macroscopic control of the harmonic content and can be readily
incorporated into superconducting circuits. Building on this approach,
we recently employed the same synthetic high-transparency element in a
transparency-engineered rf-SQUID cell for Kerr-free three-wave mixing
\cite{Guarcello2026TRAIL}.

Therefore, it is important to establish the validity of the reduction beyond the static mapping. This issue becomes particularly relevant when such a
synthetic element is employed as a building block of a driven
superconducting circuit. 
Once the two JJs are driven at finite frequency,
each junction with its own capacitive and dissipative response
introduces an internal dynamical degree of freedom with associated
time scales that are eliminated in the static CPR reduction,
as capacitive and internal dynamical effects can significantly
affect the collective behavior of coupled and series JJ systems
\cite{Filatrella1992,Chernikov1995}.
Existing rf treatments of related engineered elements have successfully
reproduced a lumped effective description in the adiabatic regime
\cite{Banszerus2025}, further supporting an effective-element description.
This question is also relevant in the broader context of nonlinear Josephson circuits, where deviations from a sinusoidal CPR can qualitatively reshape gain, stability, and nonlinear dynamical behavior
\cite{Guarcello2023TAS,Guarcello2024Chaos,Guarcello2025APL}.
A systematic comparison between the complete two-JJ dynamics and the corresponding single effective high-transparency element is therefore needed to identify the frequency and driving-amplitude ranges in which the
static mapping remains operationally valid.

In this work, we quantify the validity of the effective
high-transparency description for two JJs in series. We start from the full two-JJ model and derive an effective capacitive and dissipative
coefficients for the collective phase from a low-frequency reduction.
We then compare the purely synthetic approximation and the effective model with the complete voltage dynamics over a broad range of ac-drive frequencies, amplitudes, and junction asymmetries. 
The comparison is primarily quantified through a normalized
root-mean-square (RMS) waveform error, which probes the full time-dependent
response including amplitude, phase, and harmonic-content
differences. 
This allows us to determine the domain of validity of the
effective description and to identify the characteristic frequency
scales at which the reduction breaks down.

\section{Dynamical Model}

\begin{figure}[b]
\centering
\includegraphics[width=0.5\columnwidth]{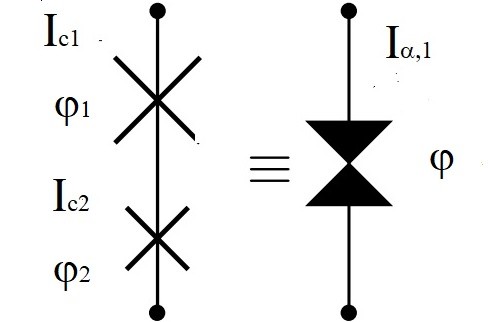}
\caption{The Josephson synthetic element: a larger JJ of critical current $I_{c1}$ in series with a smaller JJ with critical current $I_{c2}=\alpha I_{c1}$, $0 < \alpha \le 1$.}
\label{fig:circuit}
\end{figure}

We consider two conventional JJs connected in series and driven
by the same ac current (Fig.~1), with their asymmetry introduced
geometrically through the junction areas $A_1$ and $A_2$. 
By defining $\alpha \equiv A_2/A_1$ we have
\begin{equation}
\alpha
       =\frac{I_{c2}}{I_{c1}}
       =\frac{C_2}{C_1}.
\label{eq:alpha}
\end{equation}
Equation~(\ref{eq:alpha}) highlights that the ratios of the critical currents and capacitance depend only upon the ratio of the areas assuming the same critical current density and specific capacitance for both JJs, as it is natural if the two devices are fabricated on the same chip with the same procedure. 
Without loss of generality, the first JJ is the largest area junction, with critical current $I_{c1}$, capacitance $C_1$, and resistance $R_1$, and the second JJ has area $A_2 = \alpha A_1$ with $0 < \alpha \le 1$. 
The resistive terms are treated phenomenologically as subgap dissipation and are not assumed to scale with junction area.
The plasma frequency of the first larger JJ is
\begin{equation}
\omega_{p1}=\sqrt{\frac{2e I_{c1}}{\hbar C_1}}.
\end{equation}
As a consequence of Eq.~(\ref{eq:alpha}), the plasma frequencies of the two JJs are identical, for both $I_c$ and $C$ scale with junction area; thus, the two JJs have the same bare plasma frequency.
Time is normalized to the inverse of this common frequency, $\tau=\omega_{p1}t$.
The applied current reads $i(\tau)=i_{ac}\sin(\Omega\tau)$,
where currents are normalized to $I_{c1}$, voltages to $(\hbar/2e)\omega_{p1}$, so that $v=\dot{\varphi}$ and $\Omega={\omega}/{\omega_{p1}}$.

Since the JJs are connected in series, the same total current flows through both junctions, although the Josephson, capacitive, and dissipative components may differ in the two JJs.

Within the RCSJ model\cite{Stewart1968,McCumber1968,Likharev1979,Barone1982}, the dynamics of the two JJs is described by (dots denote derivatives with respect to $\tau$)
\begin{align}
\ddot{\varphi}_1+\gamma_1\dot{\varphi}_1+\sin\varphi_1
&=i(\tau),
\\
\ddot{\varphi}_2+{\gamma_2}\dot{\varphi}_2
+\sin\varphi_2
&=\frac{i(\tau)}{\alpha},
\label{eq:fullRCSJ}
\end{align}
The dissipative terms read:
\begin{equation}
\gamma_1=
\frac{\hbar\omega_{p1}}
     {2e I_{c1}R_1},
\qquad
\gamma_2=
\frac{\hbar\omega_{p1}}
     {2e I_{c2}R_2}
=
\gamma_1\frac{R_1}
     {\alpha R_2}.
\label{eq:gamma}
\end{equation}
The RCSJ framework provides the standard dynamical description of
JJs and superconducting circuits, and has also been
widely employed to investigate fluctuation-driven phenomena such as
switching and stochastic activation~\cite{Grimaudo2022,
Guarcello2023,Citro2024,DeSantis2025}.

%
%

\begin{figure*}[ht]
\centering
\includegraphics[width=0.96\textwidth]{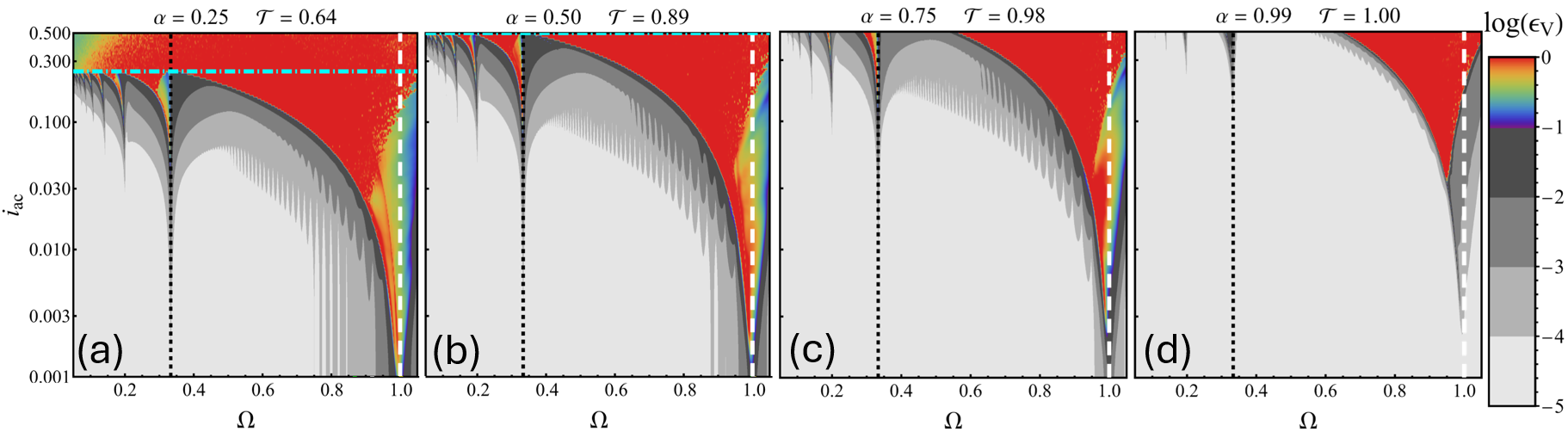}
\caption{Normalized RMS waveform error $\epsilon_V$ versus the
normalized drive frequency $\Omega$ and ac-current amplitude $i_{ac}$.
Panels (a)--(d) correspond to $\alpha=0.25$, $0.50$, $0.75$, and
$0.99$, respectively, with effective transparencies
${\cal T}\simeq\{0.64,0.89,0.98,1.00\}$.
The white dashed line marks the common plasma frequency
$\Omega_{p1}=\Omega_{p2}=\Omega_{p,{\rm eff}}=1$, while the black
dotted line indicates the third-order superharmonic condition
$\Omega=\Omega_p/3$. The cyan dot-dashed line, when within the
displayed range, marks $i_{ac}=\alpha$, corresponding to
$I_{ac}=I_{c2}$. The color scale reports $\log_{10}(\epsilon_V)$,
with values below $10^{-5}$ clipped at the numerical floor.}
\label{fig:errorMaps}
\end{figure*}

The total phase drop across the series is
$\varphi=\varphi_1+\varphi_2$, while the corresponding normalized voltage is
$v_{\rm full}=\dot{\varphi}_1+\dot{\varphi}_2$.
In the static limit, $\dot{\varphi}_i=0$ and $\ddot{\varphi}_i=0$, current conservation allows the two-JJ series element to be mapped onto a single nonsinusoidal Josephson element~\cite{Bozkurt2023,Banszerus2024,Banszerus2025}, with CPR
\begin{equation}
i_{\rm eff}(\varphi)=
\frac{\alpha\sin\varphi}
{\sqrt{1+\alpha^2+2\alpha\cos\varphi}},
\label{eq:effectiveCPR}
\end{equation}
corresponding to the effective transparency
\begin{equation}
{\cal T}=\frac{4\alpha}{(1+\alpha)^2}.
\label{eq:transparency}
\end{equation}
The effective CPR in Eq.~(\ref{eq:effectiveCPR}) is the normalized
counterpart of the synthetic-element CPR
$I_{\blacktriangleright\!\blacktriangleleft}(\varphi)$ introduced in
Ref.~\cite{Guarcello2026TRAIL}, namely $
i_{\rm eff}(\varphi)
=
{I_{\blacktriangleright\!\blacktriangleleft}(\varphi)}/{I_{c1}}.
$
We retain here the notation $i_{\rm eff}$ to distinguish this static
CPR from the different dynamical approximations introduced below.
Equation~(\ref{eq:effectiveCPR}) reproduces the functional form of the CPR of a short single-channel weak link with finite transparency~\cite{Beenakker1991,Bozkurt2023}.

To extend this mapping to finite-frequency dynamics, we associate
effective capacitive and dissipative terms with the collective phase.
In the low-frequency and small-phase limit, current conservation gives
\begin{equation}\label{eq:phases}
\varphi_1\simeq\frac{\alpha}{1+\alpha}\varphi,
\qquad
\varphi_2\simeq\frac{1}{1+\alpha}\varphi.
\end{equation}
The effective capacitive and dissipative parameters can be obtained by
requiring that, for a given collective phase $\varphi$, the electrostatic
energy stored in the effective capacitance and the power dissipated in
the effective resistive channel equal the corresponding sums over the
two individual JJs. Thus,
\begin{equation}
\frac{1}{2} C_{\rm eff}
\left(\frac{\hbar}{2e}\dot{\varphi}\right)^2
=
\frac{1}{2} C_1
\left(\frac{\hbar}{2e}\dot{\varphi}_1\right)^2
+
\frac{1}{2} C_2
\left(\frac{\hbar}{2e}\dot{\varphi}_2\right)^2,
\end{equation}
while for the dissipative contribution
\begin{equation}
\frac{1}{R_{\rm eff}}
\left(\frac{\hbar}{2e}\dot{\varphi}\right)^2
=
\frac{1}{R_1}
\left(\frac{\hbar}{2e}\dot{\varphi}_1\right)^2
+
\frac{1}{R_2}
\left(\frac{\hbar}{2e}\dot{\varphi}_2\right)^2.
\end{equation}
By time derivative of Eqs.~\eqref{eq:phases}, one obtains
\begin{equation}\label{eq:ceff}
C_{\rm eff}
=
\frac{\alpha^2 C_1+C_2}{(1+\alpha)^2},
\quad
\frac{1}{R_{\rm eff}}
=
\frac{1}{(1+\alpha)^2}
\left(
\frac{\alpha^2}{R_1}+\frac{1}{R_2}
\right)\! ,
\end{equation}
and in normalized units (since $C_2=\alpha C_1$): 
\begin{equation}
\gamma_{\rm eff}
=
\gamma_1\frac{R_1}{R_{\rm eff}}.
\qquad
c_{\rm eff}
=
\frac{C_{\rm eff}}{C_1}=\frac{\alpha}{1+\alpha}.
\label{eq:effectiveParameters}
\end{equation}
The effective dynamical model is therefore
\begin{equation}
c_{\rm eff}\ddot{\varphi}_{\rm eff}
+\gamma_{\rm eff}\dot{\varphi}_{\rm eff}
+i_{\rm eff}(\varphi_{\rm eff})
=i(\tau).
\label{eq:effectiveRCSJ}
\end{equation}
Importantly, $c_{\rm eff}$ and $\gamma_{\rm eff}$ are not fitting parameters, but follow from the low-frequency and small amplitude approximations of the original two-JJ dynamics. However, Eq.~(13) will also be employed outside these validity
limits, and the numerical analysis will quantitatively establish
its discrepancy with the full two-JJ system.

Since the plasma frequencies of the two JJs at the left of Fig.~\ref{fig:circuit} are the same, it is interesting to retrieve the effective inductance of the equivalent junction.  
The small-signal response of the effective element can be expressed
in terms of its differential Josephson inductance, consistently with
the notation adopted in Ref.~\cite{Guarcello2026TRAIL}, as
\begin{equation}
\frac{1}{L_{J,\rm eff}}
=
\frac{2\pi}{\Phi_0}
\left.
\frac{\partial I_{\rm eff}}{\partial\varphi}
\right|_{\varphi=0}
=
\frac{2\pi I_{c1}}{\Phi_0}
\frac{\alpha}{1+\alpha},
\end{equation}
where $I_{\rm eff}=I_{c1}i_{\rm eff}$.
The small-signal plasma frequency of the effective element is therefore
$\omega_{p,\rm eff}=\left(\sqrt{L_{J,\rm eff}C_{\rm eff}}\right)^{-1/2}$.
From Eq.~(\ref{eq:ceff}) the increase of the effective Josephson inductance is exactly compensated by the corresponding reduction of the effective
capacitance. Consequently,
$\omega_{p,\rm eff}=\omega_{p1}$,
or, in the normalized units adopted here,
$\Omega_{p,\rm eff}=1$.
Hence, in the present geometry,
$\Omega_{p1}=\Omega_{p2}=\Omega_{p,\rm eff}=1$.
For the numerical analysis below we further take $R_1=R_2$, so that
$\gamma_1=\alpha \gamma_2$ and
$\gamma_{\rm eff}=\gamma_1(1+\alpha^2)/(1+\alpha)^2$.

In the small-phase limit, $\sin\varphi_i\simeq\varphi_i$, and for a
harmonic drive the linearized full RCSJ equations give
\begin{equation}
\varphi_{\rm full}(\Omega)
\!=\!
\frac{i(\Omega)}
     {1-\Omega^2+\text{I}\gamma_1\Omega}
\!+\!
\frac{i(\Omega)/\alpha}
     {1-\Omega^2+\text{I}\gamma_2\Omega}
\!\simeq\!
\frac{1+\alpha}{\alpha}
\frac{i(\Omega)}{1-\Omega^2},
\label{eq:linearFull}
\end{equation}
where the last approximation holds when
$\gamma_{1,2}\Omega\ll|1-\Omega^2|$.

On the other hand, linearizing the effective CPR gives
$i_{\rm eff}(\varphi_{\rm eff})\simeq
[\alpha/(1+\alpha)]\varphi_{\rm eff}
=c_{\rm eff}\varphi_{\rm eff}$. Neglecting the corresponding
dissipative correction, Eq.~(\ref{eq:effectiveRCSJ}) therefore yields
\begin{equation}
\varphi_{\rm eff}(\Omega)
\simeq
\frac{i(\Omega)}
{c_{\rm eff}(1-\Omega^2)}
=
\frac{1+\alpha}{\alpha}
\frac{i(\Omega)}{1-\Omega^2}
\simeq
\varphi_{\rm full}(\Omega).
\label{eq:linearMatching}
\end{equation}

%
%
Thus, unlike the case of independently chosen capacitances, the effective model reproduces the linear finite-frequency response
of the series combination, apart from small dissipative corrections associated to the unequal damping coefficients.


\begin{figure*}[ht]
\centering
\includegraphics[width=0.96\textwidth]{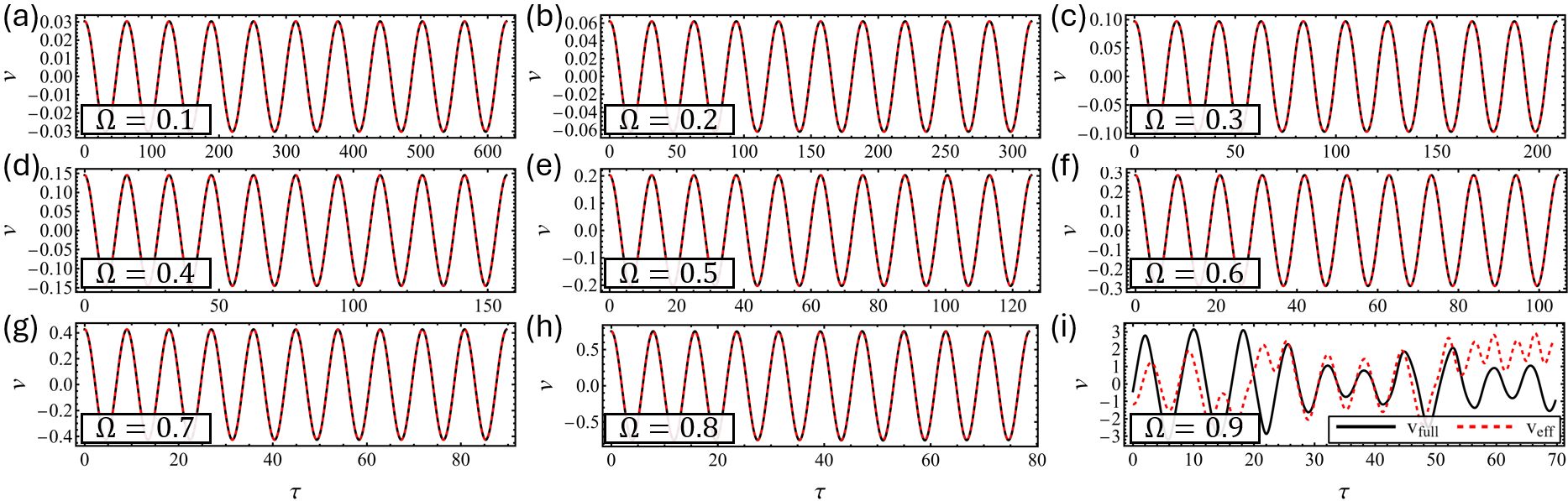}
\caption{Time-domain comparison between the normalized voltage
$v_{\rm full}$ of the complete two-JJ system (black solid line)
and $v_{\rm eff}$ of the effective model (red dashed line), for
$\alpha=0.5$ and $i_{ac}=0.1$.
Panels (a)--(i) correspond to $\Omega=0.1$, $0.2$, $0.3$, $0.4$,
$0.5$, $0.6$, $0.7$, $0.8$, and $0.9$, respectively.
The effective model closely follows the full voltage response up to
$\Omega=0.8$, including the strong increase in oscillation amplitude
on approaching resonance. A pronounced nonlinear waveform mismatch
appears at $\Omega=0.9$. The legend in panel (i) refers to all panels.}
\label{fig:waveforms}
\end{figure*}

\section{Numerical Analysis and Results}

We now compare the full two-JJ dynamics with the effective model of Eq.~(\ref{eq:effectiveRCSJ}). We take $I_{c1}=2~\mu{\rm A}$, $C_1=200~{\rm fF}$, $C_2=\alpha C_1$, $R_1=R_2=20~{k}\Omega$.
For these parameters, the plasma frequency of the reference junction is
$f_{p1}=\omega_{p1}/2\pi\simeq27.7~{\rm GHz}$ and
$\gamma_1 \simeq1.43\times10^{-3}$, with
$Q_1\simeq697$ while $\gamma_2= \gamma_1/\alpha$. Since $I_{c2}$ and $C_2$ scale by the same factor
$\alpha$, the second JJ has the same bare plasma frequency,
$\omega_{p2}=\omega_{p1}$, while remaining strongly underdamped throughout the explored range, with $Q_2=\alpha Q_1$, ranging from approximately $174$ at $\alpha=0.25$ to $690$ at $\alpha=0.99$. In fact, we explore four values of the junction asymmetry,
%
$\alpha=\{0.25,0.50,0.75,0.99\},$
%
corresponding through Eq.~(\ref{eq:transparency}) to 
${\cal T}\simeq\{0.64,0.89,0.98,1.00\}$.
For each $\alpha$, the normalized drive frequency is varied over
$0.05\leq\Omega\leq1.05$, while the ac-current
amplitude is sampled logarithmically over
$10^{-3}\leq i_{ac}\leq0.5$. No dc bias or noise
source is included.

Each pair $(\Omega,i_{ac})$ corresponds to an independent simulation
starting at the equilibrium, i.e., $\varphi_i(0)=\dot{\varphi}_i(0)=0$.
The ac drive is smoothly ramped from zero during 30 drive cycles.
The subsequent 20 cycles are discarded as transient dynamics, and the
following 50 cycles are used for the analysis.

To quantify the accuracy of the effective description, we compare the
complete voltage waveforms. The normalized voltages of the full and
effective systems are
\begin{equation}
v_{\rm full}(\tau)=
\dot{\varphi}_1(\tau)+\dot{\varphi}_2(\tau),
\qquad
v_{\rm eff}(\tau)=\dot{\varphi}_{\rm eff}(\tau),
\end{equation}
and we define the normalized RMS waveform error
\begin{equation}
\epsilon_V=
\frac{
\sqrt{\left\langle
\left[v_{\rm full}(\tau)-v_{\rm eff}(\tau)\right]^2
\right\rangle}
}{
\sqrt{\left\langle v_{\rm full}^2(\tau)\right\rangle}
}.
\label{eq:epsilonV}
\end{equation}
Here, $\langle\cdots\rangle$ denotes the time average over the drive cycles retained after the transient.
%
%
The parameter $\epsilon_V$ captures discrepancies in amplitude, phase, and
harmonic content of the complete voltage response.

%
%

This behavior is clearly reflected in Fig.~\ref{fig:errorMaps}.
For all values of $\alpha$, a broad portion of the
$(\Omega,i_{\rm ac})$--plane exhibits very small waveform errors, often
reaching $\epsilon_V\lesssim10^{-4}$--$10^{-5}$. A loss of accuracy develops when increasing $\Omega$ and $i_{\rm ac}$ produce sufficiently large phase excursions for nonlinear
and internal relative-phase dynamics to become relevant.

For the most asymmetric case, $\alpha=0.25$
[Fig.~\ref{fig:errorMaps}(a)], the high-accuracy region is already
substantial at small $i_{ac}$, but contracts as it
increases. The boundary of the high-error region bends toward lower
drive amplitudes on approaching $\Omega=1$, reflecting the resonant
enhancement of the phase response. Increasing $\alpha$ progressively
extends the domain over which the collective description remains
accurate, see Figs.~\ref{fig:errorMaps}(b) and
\ref{fig:errorMaps}(c).

The improvement becomes particularly pronounced in the nearly
symmetric case, $\alpha=0.99$ [Fig.~\ref{fig:errorMaps}(d)], where
$\epsilon_V$ remains close to the numerical floor throughout most of
the explored parameter space and appreciable deviations are confined
to the strongly nonlinear region near the plasma resonance.
This trend has a simple limiting interpretation: for $\alpha=1$,
the two normalized equations become identical and, for identical
initial conditions, $\varphi_1=\varphi_2=\varphi_{\mathrm{eff}}/2$.
The effective equation then reduces exactly to the equation of either
junction on the corresponding CPR branch. 
The effective description thus
approaches exact dynamical equivalence in the symmetric limit.

The breakdown remains strongly amplitude dependent. The cyan
dot-dashed line marks $i_{ac}=\alpha$, where the drive amplitude
reaches the smaller critical current. It provides a useful nonlinear reference scale, but does not represent a sharp
dynamical threshold. 
In addition, narrow structures visible around the black dotted line
at $\Omega\simeq1/3$ are consistent with a third-order
superharmonic resonance, $3\Omega\simeq\Omega_p$.


To illustrate directly how the discrepancy develops in the time
domain, Fig.~\ref{fig:waveforms} compares the voltage waveforms for
$\alpha=0.5$ and $i_{ac}=0.1$.
From $\Omega=0.1$ to $0.8$ [panels (a)--(h)], the two
voltage traces remain nearly indistinguishable, even though the
oscillation amplitude increases strongly as the common plasma
frequency is approached. The effective RCSJ description therefore
captures not only the adiabatic response, but also the substantial
finite-frequency dynamical enhancement occurring below resonance.

At $\Omega=0.9$ [Fig.~\ref{fig:waveforms}(i)], the response becomes
large-amplitude and strongly nonlinear, and the two trajectories
separate markedly in amplitude, phase, and waveform shape. This
loss of agreement provides the direct time-domain counterpart of the
high-error region in Fig.~\ref{fig:errorMaps}(b).

\section{Conclusions}

We have investigated the dynamical validity of replacing two conventional JJs connected in series by a single effective high-transparency Josephson element. For junctions fabricated within the same process, with $I_{c2}/I_{c1}=C_2/C_1=\alpha$, we derived effective capacitive and dissipative contributions from the static nonsinusoidal CPR mapping and compared the resulting RCSJ dynamics with the full two-JJ system.

The area scaling of both critical current and capacitance makes the two junctions and the effective element share the same plasma frequency, allowing the reduced model to reproduce the linear finite-frequency response of the full system to leading order in weak dissipation. 
The reduction remains accurate over a broad frequency--amplitude
range, with waveform errors down to $10^{-4}$--$10^{-5}$, and
improves strongly as $\alpha\to1$. Deviations mainly occur in the
strongly nonlinear near-resonant regime, where internal relative-phase
dynamics becomes relevant, thus establishing the validity range of the
single-element description for finite-frequency superconducting circuits.

\section*{Acknowledgment}

S. Pagano and C. Barone acknowledge S. Abate from CNR-SPIN Salerno for technical support.



\end{document}